\documentclass[a4paper,twocolumn,10pt]{IEEEtran}%
\usepackage{amsfonts}
\usepackage{amsmath}
\usepackage{amssymb}
\usepackage{graphicx}%
\begin{document}

\title{On the ISAR Image Analysis and Recovery with Unavailable or Heavily Corrupted Data}
\author{Ljubi\v{s}a Stankovi\'{c}, \textit{Fellow IEEE}\thanks{Electrical Engineering
Department, University of Montenegro, 20000 Podgorica, Montenegro, email:
ljubisa@ac.me, www.tfsa.ac.me. Submitted to the IEEE Transactions on Aerospace and Electronic Systems.}}
\maketitle

\begin{abstract}
Common ISAR radar images and signals can be reconstructed from much fewer
samples than the sampling theorem requires since they are usually sparse.
Unavailable randomly positioned samples can result from heavily corrupted
parts of the signal. Since these samples can be omitted and declared as
unavailable, the application of the compressive sensing methods in the
recovery of heavily corrupted signal and radar images is possible. A\ simple
direct method for the recovery of unavailable signal samples and the
calculation of the restored ISAR image is reviewed. An analysis of the noise
influence is performed. For fast maneuvering ISAR targets the sparsity
property is lost since the ISAR image is blurred. A nonparametric quadratic
time-frequency representations based method is used to restore the ISAR image
sparsity. However, the linear relation between the signal and the sparsity
domain transformation is lost. A recently proposed gradient recovery algorithm
is adapted for this kind of analysis. It does not require the linear relation
of the signal and its sparsity domain transformation in the process of
unavailable data recovery. The presented methods and results are tested on
several numerical examples proving the expected accuracy and improvements.

\begin{keywords}
Radar imaging, ISAR, time-frequency analysis, noisy signal, sparse signal,
compressive sensing.

\end{keywords}
\end{abstract}

\section{Introduction}

In inverse synthetic aperture radar (ISAR) a high resolution image of a target
is obtained by using the two-dimensional Fourier transform of the received
(and processed) signal. The ISAR image of a point target is a highly
concentrated two-dimensional pulse function at a point whose position
corresponds to the target's range and cross-range. For a number of reflecting
points, the radar image consists of several pulses at the range and
cross-positions \cite{CV}-\cite{Martorella}. Usually the number (area) of
nonzero values in the ISAR image is small as compared to the total number of
signal samples. Thus, we may say that the common signal in ISAR is sparse in
the two-dimensional Fourier domain \cite{Donoho, Candes}. As such it can be
reconstructed from much fewer samples than the sampling theorem requires.
Unavailable, randomly position samples could also result from heavily
corrupted parts of the signal, that are omitted and declared as unavailable,
before the ISAR image recovery and calculation is done \cite{LJIM}. In the
signal recovery the fact that the two-dimensional Fourier transform domain is
the domain of radar signal sparsity is used. This fact allows the application
of the compressive sensing methods \cite{Donoho, Candes,Grad0, Grad, Aut}.
A\ simple method for the unavailable radar signal data recovery and the ISAR
image calculation is reviewed in the paper. An analysis of the noise influence
on this radar image is done. A simple and accurate formula for the output
signal-to-noise ratio is derived.

For fast maneuvering ISAR targets, the radar image can be spread over the
two-dimensional Fourier transform domain \cite{CV}, \cite{Martorella2}%
-\cite{Peng}. Then a large number of the two-dimensional Fourier transform
values are nonzero, covering a large part of the radar image. In this case the
sparsity property of the signal is lost. One possibility to restore this
property is to use parametric transforms to compensate and refocus the ISAR
image, making it sparse again \cite{CV, LPFT, Peng}. However, a large number
of parameters should be used for almost each reflecting point in the case of a
general nonuniform motion. Good results can be achieved using these
techniques, but at the expense of a high computational load. This kind of
parametric calculation is even more complex for the reduced set of available
signal samples, when the compressive sensing methods are going to be used. The
other way to refocus the image is based on the quadratic time-frequency
representations \cite{CV, Compensation}. A representation which can achieve
high concentration, like in the Wigner distribution case, at the same time
avoiding the cross-terms, is the S-method. This method is nonparametric and
computationally quite simple. It requires just a few additional additions and
multiplications on the already calculated \ ISAR image using the
two-dimensional Fourier transform \cite{LJ,SM,LN}. However, the S-method
relation to the signal is not linear. Therefore, many conventional compressive
sensing based recovery techniques, including the one reviewed in this paper,
can not be used. They are based on the direct linear reconstruction relation
between the signal and the transform in the domain of signal sparsity. This is
not the case in the quadratic signal representations, such as the S-method. It
was the reason why the recently proposed gradient method for the signal
samples recovery \cite{Grad} is adapted for the problem formulation in this
paper. This method does not require a direct linear relation of the signal and
its sparsity transformation domain in the process of recovery of unavailable
signal values.

The presented methods and results are illustrated and tested on several
numerical examples proving the expected efficiency and improvements.

The manuscript is organized as follows. A brief review of the signal model in
the considered ISAR systems is given in Section 2. A reconstruction algorithm
for the radar signal with unavailable data is presented in Section 3, along
with the analysis of noise influence. The gradient method for the
reconstruction of the ISAR images, corresponding to nonuniform motion is
presented in Section 4. Examples illustrate the accuracy of the proposed methods.

\section{Radar Signal Model}

\begin{figure}
[ptb]
\begin{center}
\includegraphics[
height=1.3889in,
width=3.183in
]%
{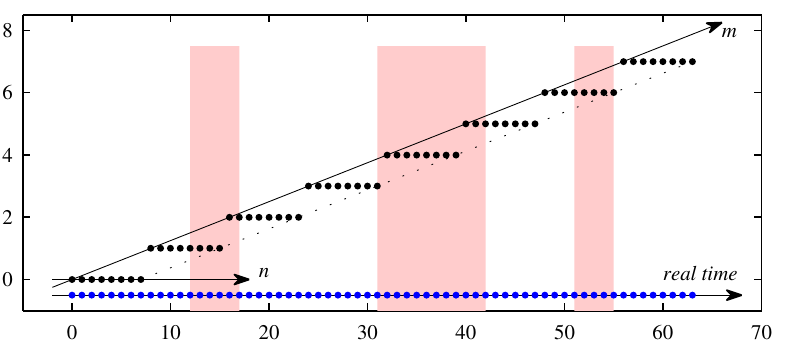}%
\caption{Illustration of one revisit (chirp series) discretization in
coordinates $m$ (chirp index, slow time) and $n$ (time within one chirp, fast
time), along with a real time. The case of $M=8$ chirps in one revisit and
$N=8$ samples within chirp is presented. The CIT is $64$ samples. Unavailable
or heavily corrupted data are marked by read.\ }%
\label{chdisk_ill}%
\end{center}
\end{figure}

For a continuous wave radar that transmits signal in a form of series of $M$
chirps \cite{CV} the received signal (reflected from a target) is delayed with
respect to the transmitted signal for $t_{d}=2d(t)/c$, where $d(t)$ is the
target distance from the radar and $c$ is the speed of light. The received
signal, after an appropriate demodulation, compensation and filtering, is%
\begin{equation}
q(m,t)=\sigma e^{j\Omega_{0}\frac{2d}{c}}e^{-j2\pi Bf_{r}(t-mT_{r})\frac
{2d}{c}}%
\end{equation}
where $\sigma$ is the reflection coefficient of the target, while $\Omega_{0}$
is the radar operating frequency. The repetition time of a single chirp is
denoted by $T_{r}$, while the\ number of samples within each chirp is $N$. The
coherent integration time (CIT) is $T_{c}=MT_{r}$. Index $m$ corresponds to
the chirp index (slow time). The received signal for a system of point
scatterers can be modeled as a sum of the individual point scatterer
responses, \cite{CV}. The Doppler part in the received signal of a point
target is
\begin{equation}
s(t)=\sigma e^{j2d(t)\Omega_{0}/c}, \label{sit}%
\end{equation}
By denoting $t-mT_{r}=nT_{s},$ where $T_{s}=T_{r}/N$ is a sampling interval
within a chirp and $n$ is the index of signal sample within one chirp
(fast-time), the range part of the received signal $\exp(-j2\pi Bf_{r}%
(t-mT_{r})\frac{2d}{c})$ reduces to $\exp(j2\pi\gamma n/N)$ with
$\gamma=-Bf_{r}T_{s}N(2d/c)$. The two-dimensional Fourier transform of the
received and processed signal is
\begin{equation}
Q(k,l)=\sum_{m=0}^{M-1}\sum_{n=0}^{N-1}q(m,n)\exp e^{-j(\frac{2\pi mk}%
{M}+\frac{2\pi nl}{N})},
\end{equation}
The illustration of the discrete $q(m,n)$ values in one revisit is presented
in Fig.~\ref{chdisk_ill}.%

\subsection{Uniform Target Motion with Unavailable/Corrupted Data}

In the simplest case, when the target motion may be considered as uniform
within the CIT, the distance can be written as
\[
d(t)\cong d_{0}+vt\cong d_{0}+vmT_{r}.
\]
The received signal, from the $i$th reflecting point, after the distance
compensation, is
\begin{align*}
q_{i}(m,t)  &  =\sigma_{i}e^{j\Omega_{0}2v_{i}T_{r}m/c}e^{j2\pi\gamma_{i}%
n/N}\\
&  =\sigma_{i}e^{j2\pi\beta_{i}m/M}e^{j2\pi\gamma_{i}n/N},
\end{align*}
where $\beta_{i}$ and $\gamma_{i}$ are the constants proportional to the
velocity (cross-range) and range. The total signal for $K$ reflecting points
is
\[
q(m,n)=\sum_{i=1}^{K}q_{i}(m,n).
\]
Next assume that some blocks of the received radar signal are either
unavailable or highly corrupted so that it is better to omit them from the
analysis \cite{LJIM}. Assume that the blocks of the omitted signal samples are
randomly positioned. The two-dimensional Fourier transform of this signal is
then
\[
\hat{Q}(k,l)=\sum_{m=0}^{M-1}\sum_{n\in\mathbf{N}_{A}(m)}q(m,n)e^{-j(\frac
{2\pi mk}{M}+\frac{2\pi nl}{N})}.
\]
It can happen that the unavailable/corrupted data are: all within one chirp or
spread over two or more chirps, including the possibility that a few chirps in
a row are affected in this way, Fig.~\ref{chdisk_ill}. These cases are
included by using the notation $n\in\mathbf{N}_{A}(m)$ where $\mathbf{N}%
_{A}(m)$ is the set of available samples within the $m$th chirp. \ For some
$m$ it could also happen that $\mathbf{N}_{A}(m)=\varnothing$, i.e., that
there are no available samples within that chirp. The total number of
available samples is $1\ll N_{A}\le MN$. We can consider two cases
\cite{Stan11}:

(1) For $k=\beta_{i}$ and $l=\gamma_{i}$ we will have
\begin{equation}
\hat{Q}(k,l)=\sum_{m=0}^{M-1}\sum_{n\in\mathbf{N}_{A}(m)}\sigma_{i}=\sigma
_{i}N_{A} \label{Na_EST}%
\end{equation}
where $N_{A}$ is the total number of available samples.

(2) For $k\neq\beta_{i}$ or $l\neq\gamma_{i}$ then%
\[
\hat{Q}(k,l)=\sum_{m=0}^{M-1}\sum_{n\in\mathbf{N}_{A}(m)}\sigma_{i}%
e^{j\phi(n,m,k,l)}=\Xi(k,l).
\]
For a large number of unavailable samples $1\ll N_{A}\ll NM$ the previous
value is a sum or vectors with quasi arbitrary phases. It can be considered as
a complex-valued variable with Gaussian distributed real and imaginary parts
(as shown in \cite{LJSSA}). Its variance is
\[
\mathrm{var}\{\hat{Q}(k,l)\}=N_{A}\frac{NM-N_{A}}{NM-1}\sigma_{i}^{2}.
\]
Therefore, for $K$ reflecting points we may write \cite{LJSSA}
\begin{align*}
\mathrm{E}\{\hat{Q}(k,l)\}  &  =\sum_{i=1}^{K}\sigma_{i}N_{A}\delta
(k-\beta_{i},l-\gamma_{i})\\
\mathrm{var}\{\hat{Q}(k,l)\}  &  =N_{A}\frac{NM-N_{A}}{NM-1}\sum_{i=1}%
^{K}\sigma_{i}^{2}\left(  1-\delta(k-\beta_{i},l-\gamma_{i})\right)  .
\end{align*}
Based on this analysis, the received signal and the ISAR image recovery can be
done using the following simple and computationally efficient algorithm.

\smallskip

\textbf{Algorithm:}

(i) Calculate the initial transform estimate $\hat{Q}(k,l)$ by using the
available/remaining signal values
\begin{gather}
\hat{Q}(k,l)=\sum_{m=0}^{M-1}\sum_{n\in\mathbf{N}_{A}(m)}q(m,n)e^{-j(\frac
{2\pi mk}{M}+\frac{2\pi nl}{N})}.\label{MS_SUM}\\
\text{ or \ \ \ \ }\mathbf{\hat{Q}}\mathbf{=\Phi y}\text{\textbf{.}}\nonumber
\end{gather}
where $\mathbf{y}$ is the vector of available samples $q(m,n)$, $n\in
\mathbf{N}_{A}(m)$%
\[
\mathbf{y=[}q(m,n)\left\vert \text{ \ }n\in\mathbf{N}_{A}(m)\right.  ]^{T}.
\]
Note that the two-dimensional data $q(n,m)$ are transformed into a column
vector $\mathbf{y}$ and $\mathbf{\Phi}$ is the corresponding transformation
matrix. It is used to produce $\hat{Q}(k,l)$ arranged into a column vector
$\mathbf{\hat{Q}}$.

(ii) Set the resulting transform values $Q(k,l)$ to zero at all positions
$(k_{i},l_{i})$ except the highest $\hat{K}$ values in the initial estimate
$\hat{Q}(k,l)$, i.e.,
\begin{align*}
Q(k,l)  &  =0~~\text{for }(k,l)\neq(k_{i},l_{i})\text{, }i=1,2,...,\hat{K}\\
(k_{i},l_{i})  &  =\underset{i=1,2,...,\hat{K}}{\arg}\{\max\{\mathrm{sort}%
\{\left\vert Q(k,l)\right\vert \}\}\}.
\end{align*}
This criterion is not sensitive to the assumed number of nonzero coefficients
$\hat{K}$ as far as all nonzero positions of the original transform are
detected and the total number $\hat{K}$ of transform values in $\hat{Q}(k,l)$
is lower than the number of available samples, i.e.,
\[
K\leq\hat{K}\leq N_{A}.
\]
All $\hat{K}-K$ transform values that a zero in the original signal will be
found as zero-valued in the algorithm.

(iii) The unknown $\hat{K}$ transform coefficients could be then easily
calculated by solving the set of $N_{A}$ equations for available instants
$n\in\mathbf{N}_{A}(m)$, at the detected nonzero candidate positions
$(k_{i},l_{i})$, $i=1,2,...,\hat{K}$. The linear system for unknowns
$Q(k_{i},l_{i})$ is obtained using the inverse two-dimensional Fourier
transform for $N_{A}$ available signal values,
\begin{gather}
\frac{1}{MN}%
{\textstyle\sum\nolimits_{i=1}^{\hat{K}}}
Q(k_{i},l_{i})e^{j(\frac{2\pi mk_{i}}{M}+\frac{2\pi nl_{i}}{N})}=q(m,n),\text{
}\label{Sig_REC0}\\
\text{for }0\leq m\leq N-1\text{, \ \ }n\in\mathbf{N}_{A}(m)\text{. }\nonumber
\end{gather}
System (\ref{Sig_REC0}) is a system of $N_{A}$ linear equations with $\hat{K}$
unknown transform values $Q(k_{i},l_{i})$. This linear system can be written
in a matrix form as
\[
\mathbf{\Psi Q}_{\hat{K}}\mathbf{=y,}%
\]
where: $\mathbf{Q}_{\hat{K}}$ is a vector whose elements are unknowns
$Q(k_{i},l_{i})$, $i=1,2,...,\hat{K}$, $\mathbf{\Psi}$ is the corresponding
coefficients matrix, and $\mathbf{y}$ is a vector whose elements are available
signal $q(m,n)$ samples.

For $\hat{K}=N_{A}$ its solution is simple, $\mathbf{Q}_{K}=\mathbf{\Psi}%
^{-1}\mathbf{y}$. In general, for $\hat{K}<N_{A}$ the system is solved in the
least square sense as
\begin{equation}
\mathbf{Q}_{K}=\left(  \mathbf{\Psi}^{H}\mathbf{\Psi}\right)  ^{-1}%
\mathbf{\Psi}^{H}\mathbf{y.} \label{Sig_REC}%
\end{equation}
where $H$ denotes the Hermitian transpose operation. The reconstructed
coefficients $Q(k_{i},l_{i})$, $i=1,2,...,\hat{K}$, (vector $\mathbf{Q}%
_{\hat{K}}$) are equal to the transform coefficients of the original signal
for all detected candidate frequencies. If some transform coefficients, whose
true value should be zero, are included (when $K<\hat{K}$) the resulting
system will produce their correct (zero) values.

The condition that the system (\ref{Sig_REC0}), with $\hat{K}$ unknowns, has a
solution is that there are at least $\hat{K}$ independent equations, i.e.,
that
\begin{gather*}
\mathrm{{rank}}(\Psi)\geq\hat{K}\text{ \ or}\\
\det(\mathbf{\Psi}^{H}\mathbf{\Psi)}\mathbf{\neq}0\mathbf{.}%
\end{gather*}

The reconstruction accuracy can be easily checked calculating the mean squared
error between the reconstructed samples and the available samples, at the
positions of the available samples $n\in\mathbf{N}_{A}(m)$.\medskip

\textbf{Comments:}

In general, a simple strategy can be used by assuming that $\hat{K}=N_{A}$ and
by setting to zero-value the smallest $N-N_{A}$ transform coefficients in
$\hat{Q}(k,l)$. This simple strategy is very efficient if there is no input
noise. Large $\hat{K}$, close or equal to $N_{A}$, will increase the
probability that full signal recovery is achieved in one step. However, in the
case of additional input noise in available samples, a value of $\hat{K}$ as
close to the true signal sparsity $K$ as possible will reduce the noise
influence on the reconstructed signal. This will be shown later.

If the algorithm fails to detect a component (the reconstruction accuracy of
the available samples can be used to detect this event) the procedure can be
repeated after the detected components are reconstructed and removed. In such
cases the iterative procedure is recommended.\medskip

\textbf{Iterative procedure: }

If the number of available samples is low or there are components with much
lower amplitudes, so that they can not be detected in one step, the iterative
procedure should be used. Algorithm for the iterative procedure is:

\noindent-The largest component at $(k_{1},l_{1})$ in (\ref{MS_SUM}) is
detected. The transform values $Q(k,l)$ are set to zero at all positions
$(k,l)$ except at the position of the highest one at $(k_{1},l_{1})$. This
component is reconstructed using (\ref{Sig_REC0}) with $\hat{K}=1$ and
subtracted from the signal.

\noindent-The remaining signal is used to calculate (\ref{MS_SUM}) again. The
highest value position $(k_{2},l_{2})$ is found, and signal is reconstructed
at two frequency points $\{(k_{1},l_{1}),(k_{2},l_{2})\}$ using
(\ref{Sig_REC0}) with $\hat{K}=2$. The reconstructed signal is removed from
the original signal and (\ref{MS_SUM}) is calculated with the remaining signal.

\noindent-Procedure is continued in this way until the maximal absolute
difference of the reconstructed signal, with $\hat{K}$ components at positions
$\{(k_{1},l_{1}),(k_{2},l_{2}),...,(k_{\hat{K}},l_{\hat{K}})\}$, and the given
available signal values (at the positions $n\in\mathbf{N}_{A}(m)$) is bellow
the required accuracy level. \medskip

\begin{figure}
[ptb]
\begin{center}
\includegraphics[
height=2.9058in,
width=2.9365in
]%
{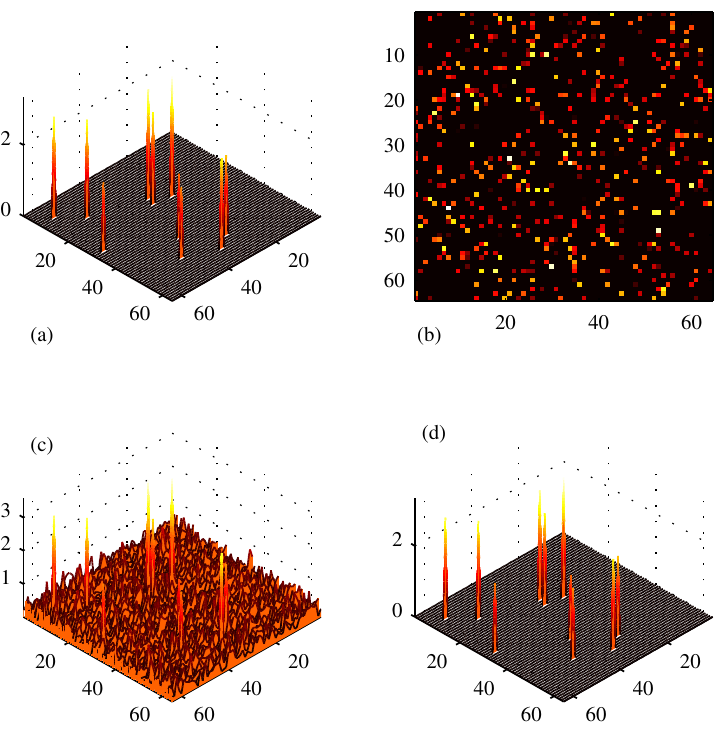}%
\caption{(a) A two-dimensional Fourier transform of the considered radar
signal (ISAR image). (b) Radar signal with 12.5\% of available/uncorrupted
samples (unavailable/corrupted samples are presented in black). (c) The
two-dimensional Fourier transform calculated using the available samples of
the radar signal. (d) The recontracted ISAR image.}%
\label{dft2d_omp}%
\end{center}
\end{figure}

\noindent\textbf{\textit{{Example 1:}}} A signal with $K=10$ randomly
positioned reflecting points%
\[
q(m,n)=\sum_{i=1}^{10}\sigma_{i}e^{j2\pi\beta_{i}m/M}e^{j2\pi\gamma_{i}n/N},
\]
with reflecting coefficients $1/8\leq\sigma_{i}\leq3/8$ and $M=N=64$ is
considered with $87.5\%$ unavailable samples. The two-dimensional Fourier
transform (ISAR image) of the original signal, if all signal samples were
available, is presented in Fig.\ref{dft2d_omp}(a). The initial two-dimensional
Fourier transform of the signal is calculated using (\ref{MS_SUM}) with
$N_{A}=0.125MN$ available samples, Fig.\ref{dft2d_omp}(b). It is presented in
Fig.\ref{dft2d_omp}(c). The largest $\hat{K}=14>10$ values in $\hat{Q}(k,l)$
are taken as candidates for the nonzero coefficients. Note that any
$10\leq\hat{K}\leq512$ would produce the same result, as far as the nonzero
coefficients of the original signal's two-dimensional Fourier transform are
included. The signal is then fully reconstructed using (\ref{Sig_REC0}%
)-(\ref{Sig_REC}) and presented in Fig.\ref{dft2d_omp}(d). The difference
between the available signal values and the reconstructed signal values, at
the same positions, is within the computer precision.%

\subsection{Influence of Additive Input Noise}

Assume now that an input additive noise $\varepsilon(n)$ exists in the
available data. Note that the noise due to missing values influences the
results in the sense presented in the previous section. When the recovery is
achieved the accuracy of the result is related to the input additive noise in
signal samples. It also depends on the number of available signal samples and
nonzero transform coefficients (sparsity) as it will be shown next.

The reconstruction equations (\ref{Sig_REC0}) for the noisy available data
are
\begin{align}
q(m,n)+\varepsilon(m,n)  &  =\frac{1}{MN}%
{\textstyle\sum\nolimits_{i=1}^{\hat{K}}}
Q(k_{i},l_{i})e^{j(\frac{2\pi mk_{i}}{M}+\frac{2\pi nl_{i}}{N})}%
,\label{Noise_rec}\\
\text{for }0  &  \leq m\leq N-1\text{, \ \ }n\in\mathbf{N}_{A}(m)\text{.}%
\nonumber
\end{align}
The transform indices can take a value from the set of detected values
$(k,l)\in\{(k_{1},l_{1}),(k_{2},l_{2}),...,(k_{\hat{K}},l_{\hat{K}})\}$. A
matrix form of equations (\ref{Noise_rec}) is%

\[
\mathbf{y+\varepsilon}=\mathbf{\Psi Q}_{\hat{K}}.
\]
This is a system of $N_{A}$ linear equations with $\hat{K}$ unknowns in
$\mathbf{Q}_{\hat{K}}$. As it has been shown, the solution is%
\begin{align}
\mathbf{\Psi}^{H}(\mathbf{y+\varepsilon})  &  =\mathbf{\Psi}^{H}\mathbf{\Psi
Q}_{\hat{K}}\nonumber\\
\mathbf{Q}_{\hat{K}}  &  =\left(  \mathbf{\Psi}^{H}\mathbf{\Psi}\right)
^{-1\mathbf{\Psi}H}(\mathbf{y+\varepsilon})\nonumber\\
\mathbf{Q}_{\hat{K}}  &  =\mathbf{Q}_{KS}+\mathbf{Q}_{KN}. \label{REC_ALG}%
\end{align}
The true transform coefficients and the noise influence to the reconstructed
transform are
\begin{align*}
\mathbf{Q}_{KS}  &  =\left(  \mathbf{\Psi}^{H}\mathbf{\Psi}\right)
^{-1}\mathbf{\Psi}^{H}\mathbf{y,}\\
\mathbf{Q}_{KN}  &  =\left(  \mathbf{\Psi}^{H}\mathbf{\Psi}\right)
^{-1}\mathbf{\Psi}^{H}\mathbf{\varepsilon.}%
\end{align*}

If all signal samples were available, the input signal-to-noise (SNR) ratio,
would be%

\begin{equation}
SNR_{i}=10\log\frac{\sum_{m=0}^{M-1}\sum_{n=0}^{N-1}\left\vert
q(m,n)\right\vert ^{2}}{\sum_{m=0}^{M-1}\sum_{n=0}^{N-1}\left\vert
\varepsilon(m,n)\right\vert ^{2}}=10\log\frac{E_{s}}{E_{\varepsilon}}.
\label{SNRi}%
\end{equation}
Assume that the noise energy in the available samples is%
\begin{equation}
E_{\varepsilon A}=\sum_{m=0}^{M-1}\sum_{n\in\mathbf{N}_{A}}\left\vert
\varepsilon(m,n)\right\vert ^{2}.
\end{equation}
The true amplitude in the signal transform at the position $(k_{i},l_{i})$, in
the case if all signal samples were used, would be $MN\sigma_{i}$ where
$\sigma_{i}$ is the amplitude (reflection coefficient) of the signal component
corresponding to the position $(k_{i},l_{i})$. To compensate the resulting
transform for the known bias in amplitude (\ref{Na_EST}) when only $N_{A}$
available samples are used the coefficient should be multiplied by $MN/N_{A}$.
In a full recovery, a signal transform coefficient is equal to the coefficient
of the original signal with all signal samples being used. The noise in the
transform coefficients is multiplied by the same factor of $MN/N_{A}$.
Therefore, the energy of noise in the reconstruction algorithm is increased to
$E_{\varepsilon A}(MN/N_{A})^{2}$. The SNR in the recovered signal is
\begin{equation}
SNR=10\log\frac{\sum_{m=0}^{M-1}\sum_{n=0}^{N-1}\left\vert q(m,n)\right\vert
^{2}}{\frac{M^{2}N^{2}}{N_{A}^{2}}\sum_{m=0}^{M-1}\sum_{n\in\mathbf{N}_{A}%
}\left\vert \varepsilon(m,n)\right\vert ^{2}}%
\end{equation}

Since only $\hat{K}$ out of $MN$ coefficients are used in the reconstruction
the energy of the reconstruction error is reduced for the factor of $\hat
{K}/(MN)$ as well. The energy of noise in the recovered signal is
\[
E_{\varepsilon R}=\frac{\hat{K}}{MN}\frac{M^{2}N^{2}}{N_{A}^{2}}\sum
_{m=0}^{M-1}\sum_{n\in\mathbf{N}_{A}}\left\vert \varepsilon(m,n)\right\vert
^{2}.
\]
The SNR in the recovered signal is%
\begin{equation}
SNR=10\log\frac{\sum_{m=0}^{M-1}\sum_{n=0}^{N-1}\left\vert q(m,n)\right\vert
^{2}}{\frac{\hat{K}NM}{N_{A}^{2}}\sum_{m=0}^{M-1}\sum_{n\in\mathbf{N}_{A}%
}\left\vert \varepsilon(m,n)\right\vert ^{2}}. \label{SNR_RECOV}%
\end{equation}
Since the variances of noise in all samples and the available samples are the
same then
\begin{equation}
\frac{1}{N_{A}}\sum_{m=0}^{M-1}\sum_{n\in\mathbf{N}_{A}}\left\vert
\varepsilon(m,n)\right\vert ^{2}=\frac{1}{MN}\sum_{m=0}^{M-1}\sum_{n=0}%
^{N-1}\left\vert \varepsilon(m,n)\right\vert ^{2} \label{CQ}%
\end{equation}
Thus, the SNR in the recovered signal, according to (\ref{SNR_RECOV}),
(\ref{CQ}) and (\ref{SNRi}), is
\begin{equation}
SNR=SNR_{i}-10\log\left(  \frac{\hat{K}}{N_{A}}\right)  . \label{Theor_SNR}%
\end{equation}

We may conclude that in the case of additive input noise in the available
signal samples, the output SNR will be increased if the number $\hat{K}$ is as
small as possible, for a given number of available samples $N_{A}$. In the
ideal case, with respect to the additive noise, value of $\hat{K}$ should be
equal to the signal sparsity $\hat{K}=K$.

\medskip

\noindent\textbf{\textit{{Example 2:}}} Consider a noisy signal from Example
1. Assume that an additive complex-valued Gaussian noise exists, with the
input SNR equal to
\[
SNR_{i}=9.05\text{ [dB]}%
\]
and $N_{A}=MN/8$. Since $K=10$ in the previous example we used estimated value
$\hat{K}=14$ for the calculation. According to (\ref{Theor_SNR}) the output
SNR is
\begin{align*}
SNR  &  =SNR_{i}-10\log\left(  \frac{\hat{K}}{N_{A}}\right) \\
&  =9.05+15.81=24.86\text{ [dB].}%
\end{align*}
The improvement in SNR is $15.81$ [dB]. This result is statistically checked.
The statistical result is obtained by averaging over 100 realizations. The
obtained statistical value of the output SNR is%
\[
SNR^{(stat)}=24.53\text{ [dB].}%
\]
Agreement with the theory is almost exact, within the number of realizations
statistical confidence.

If the number of components was estimated exactly as $\hat{K}=10,$ then the
SNR values would be obtained as
\begin{align*}
SNR  &  =26.32\text{ [dB]}\\
SNR^{(stat)}  &  =26.26\text{ [dB].}%
\end{align*}
The SNR value for $\hat{K}=10$ would be higher for $10\log(14/10)=1.46$ [dB]
than in the case with $\hat{K}=14$.

\subsection{Nonuniform Target Motion}

For fast moving targets and complex motions, the target over all $M$ chirps,
in one revisit, cannot be considered as the one with constant velocity motion.
Then a higher-order approximation%
\[
d(t)\cong d_{0}+v_{0}t+a\frac{t^{2}}{2}+\dots,
\]
should be used with%
\[
v(t)=v_{0}+at+\dots
\]
If we assume that $v(t)=v_{0}+at$, then the Doppler shift is linear function
of time. Its rate is $a$. Thus, instead of a delta pulse concentrated at one
frequency, corresponding to $v_{0}$, we will obtain a Fourier transform of a
linear frequency-modulated signal (or higher-order frequency-modulated
signal), whose instantaneous frequency changes are proportional to the
velocity $v(t)$ changes. The radar image, based on this form, is centered at
the same position as the Fourier transform image, but with the spreading term
in the cross-range (Doppler) direction of the form $\exp\left(  j\frac
{2\Omega_{0}}{c}(\frac{1}{2!}d^{\prime\prime}(0)t^{2}+...)\right)  $, due to
the target motion. In the discrete domain the signal is \cite{LJ}
\begin{align*}
q_{i}(m,n)  &  =\sigma_{i}e^{j2\pi\beta_{i}m/M}e^{j\alpha_{i}m^{2}%
/2+...}e^{j2\pi\gamma_{i}n/N},\\
Q_{i}(k,l)  &  =\left(  2\pi\right)  ^{2}\sigma_{i}\delta(k-\beta_{i}%
,l-\gamma_{i})\ast_{k}\mathrm{FT}\{e^{j\alpha_{i}m^{2}/2+...}\}
\end{align*}
where $\alpha_{i}=2\Omega_{0}T_{r}^{2}d^{\prime\prime}(0)/c$ and $\ast_{k}$ is
the convolution in the discrete cross-range domain. \ This spread can be
significant and the resulting ISAR image is not sparse or sparsity is
significantly degraded.

If the two-dimensional Fourier transform is corrected according to the
S-method \cite{LJ, SM, LN}, along the cross-range direction, then the
resulting image will be
\[
SM_{i}(k,l)=\left(  2\pi\right)  ^{2}\sigma_{i}^{2}\delta(k-\beta_{i}%
,l-\gamma_{i}).
\]
It is sparse again and does not depend on $d^{\prime\prime}(0)$. Under certain
conditions this representation is free of cross-terms among different
reflection points, producing
\[
SM(k,l)=\left(  2\pi\right)  ^{2}\sum\limits_{i=1}^{K}\sigma_{i}^{2}%
\delta(k-\beta_{i},l-\gamma_{i}).
\]

The S-method based ISAR image can be easily realized in a recursive way
starting from
\begin{equation}
SM_{0}(k,l)=\left\vert Q(k,l)\right\vert ^{2},
\end{equation}
with $SM_{0}(k)$ being the standard two-dimensional Fourier transform based
radar image. The S-method based presentation can be achieved starting with the
already obtained Fourier transform-based radar image $Q(k,l)$, with an
additional simple calculation according to%
\[
SM_{L}(k,l)=SM_{L-1}(k,l)+2\operatorname{Re}\{Q(k+L,l)Q^{\ast}(k-L,l)\}
\]
or
\begin{gather}
\mathrm{SM}_{L}[q(m,n)]=SM_{L}(k,l)\label{SM_CORR}\\
=\left\vert Q(k,l)\right\vert ^{2}+2\sum_{z=1}^{L}\operatorname{Re}%
\{Q(k+z,l)Q^{\ast}(k-z,l)\}\nonumber
\end{gather}

In this way, using the S-method, we will restore signal sparsity in the ISAR
image domain. However we have lost the possibility to use a direct linear
relation between the signal and the sparsity domain transformation. For a
reduced set of $N_{A}<MN$ available signal samples, $n\in\mathbf{N}_{A}(m)$
the problem statement is now
\begin{equation}
\min\left\Vert SM_{L}(k,l)\right\Vert _{0}\text{ subject to the available
values }\mathbf{y}. \label{DEF_L0}%
\end{equation}
where $\mathbf{y}$ is the vector of the available signal samples $q(m,n)$,
$n\in\mathbf{N}_{A}(m)$ and
\[
\left\Vert SM_{L}(k,l)\right\Vert _{0}=\sum\limits_{k=0}^{N-1}\sum
\limits_{l=0}^{N-1}\left\vert SM_{L}(k,l)\right\vert ^{0}.
\]
The simple counting of nonzero coefficients by using the zero-norm with
$\left\vert SM(k,l)\right\vert ^{0}$ is, in theory, the best optimization
function. Finding the unavailable signal value to produce the minimal number
of nonzero coefficients in the resulting presentation is an obvious
optimization criterion. However, this criterion is very sensitive to small
values in $SM_{L}(k,l)$. Also the gradient solutions are not possible with the
zero-norm functions, since they are completely flat for any nonoptimal value.
That is why the norm-one is used in the standard compressive sensing methods
instead of the norm-zero. In the S-method formulation the norm that will
correspond to the commonly used norm-one of the Fourier transform is
\begin{equation}
\min\left\Vert SM_{L}(k,l)\right\Vert _{1/2}\text{ subject to available values
of }\mathbf{y}. \label{DEF_L1}%
\end{equation}
with
\[
\left\Vert SM_{L}(k,l)\right\Vert _{1/2}=\sum\limits_{k=0}^{N-1}%
\sum\limits_{l=0}^{N-1}\left\vert SM_{L}(k,l)\right\vert ^{1/2}.
\]
This form, for $L=0$ reduces to the norm-one of the Fourier transform, since
\[
\sum\limits_{k=0}^{N-1}\sum\limits_{l=0}^{N-1}\left\vert SM_{0}%
(k,l)\right\vert ^{1/2}=\sum\limits_{k=0}^{N-1}\sum\limits_{l=0}%
^{N-1}\left\vert Q(k,l)\right\vert =\left\Vert Q(k,l)\right\Vert _{1}%
\]
Minimization of $\sum\nolimits_{k=0}^{N-1}\sum\nolimits_{l=0}^{N-1}\left\vert
SM_{L}(k,l)\right\vert ^{1/2}$ has already been used for the time-frequency
parameters optimization in \cite{Ljubmeasure}. Note that under certain
conditions the norm-one produces the same result as the norm-zero in the
problem formulation (\ref{DEF_L0}), \cite{Donoho, Candes}.

A simple gradient algorithm to iteratively calculate the missing signal
values, while keeping available samples $q(m,n)$ unchanged, \cite{Grad}, is
adapted for the problem formulation (\ref{DEF_L1}). It is presented next.

\subsection{Algorithm}

This gradient algorithm is inspired by the adaptive signal processing methods
with an adaptive step size. It is a gradient descent algorithm where the
missing samples, are corrected according to the gradient of the sparsity
measure $\left\Vert SM_{L}(k,l)\right\Vert _{1/2}$. Their final value should
converge to the point of the minimal sparsity measure of the signal
time-frequency representation.

The algorithm for missing samples reconstruction is implemented as follows:

\bigskip

\noindent\textbf{Step 0:} Set $s=0,$ $p=0$ and form the initial signal
$y^{(0)}(m,n)$ defined for all $\ m$ and $n$ as:
\[
y^{(0)}(m,n)=\left\{
\begin{array}
[c]{ll}%
q(m,n) & \text{for available samples, }n\in\mathbf{N}_{A}(m)\\
0 & \text{for }n\notin\mathbf{N}_{A}(m)
\end{array}
\right.  ,
\]
The initial value for an algorithm parameter $\Delta$ is estimated as%
\begin{equation}
\Delta=\max_{n\in\mathbf{N}_{A}(m)}|q(m,n)|. \label{deltaINIt}%
\end{equation}

\noindent\textbf{Step 1:} Set $y_{r}(m,n)=y^{(p)}(m,n)$. This signal is used in Step 3
in order to estimate reconstruction precision.\newline\textbf{Step 2.1:} Set
$p=p+1$. For each missing sample at $(n_{i},m_{i})$ for $n\notin\mathbf{N}%
_{A}(m)$ form the signals $y_{1}(m,n)$ and $y_{2}(m,n)$:
\begin{align}
y_{1}(m,n)  &  =y^{(p)}(m,n)+\Delta\delta(n-n_{i},m-m_{i})\nonumber\\
y_{2}(m,n)  &  =y^{(p)}(m,n)-\Delta\delta(n-n_{i},m-m_{i}). \label{Sig_Delta}%
\end{align}
\noindent\textbf{Step 2.2: }Estimate differential of the signal transform
measure%
\begin{equation}
g(m_{i},n_{i})=\frac{\left\Vert SM_{1,L}(k,l)\right\Vert _{1/2}-\left\Vert
SM_{2,L}(k,l)\right\Vert _{1/2}}{2N\Delta} \label{eq:mjera}%
\end{equation}
where $SM_{1,L}(k,l)=\mathrm{SM}_{L}[y_{1}(m,n)]$ and $SM_{2,L}%
(k,l)=\mathrm{SM}_{L}[y_{2}(m,n)]$ are the S-methods of $y_{1}(m,n)$ and
$y_{2}(m,n)$, respectively, calculated with $L$ correction terms.

\noindent\textbf{Step 2.3:} Form a gradient matrix $\mathbf{G}_{p}$ with the
same size as the signal $q(m,n)$. At the positions of available samples
$n\in\mathbf{N}_{A}(m)$, this vector has value $G_{p}(m,n)=0$. At the
positions of missing samples $n_{i}\notin\mathbf{N}_{A}(m)$ its values are
$G_{p}(m,n)=g(m_{i},n_{i})$, calculated by (\ref{eq:mjera}).

\noindent\textbf{Step 2.4:} Correct the values of $y(m,n)$ iteratively by
\begin{equation}
y^{(p)}(m,n)=y^{(p-1)}(m,n)-2\Delta G_{p}(m,n),
\end{equation}
\textbf{Step 3:} If the maximal allowed number of iterations $P_{\max}$ is
reached stop the algorithm. Otherwise calculate%
\[
T_{r}=\frac{\sum_{m=0}^{M-1}\sum_{n\notin\mathbf{N}_{A}}|y_{r}(m,n)-y^{(p)}%
(m,n)|^{2}}{\sum_{m=0}^{M-1}\sum_{n\notin\mathbf{N}_{A}}|y^{(p)}(m,n)|^{2}}.
\]
Value of $T_{r}$ is an estimate of the reconstruction error to signal ratio,
calculated for missing samples only. If $T_{r}$ is above the required
precision threshold (for example, if $T_{r}>0.001$), the calculation procedure
should be repeated with smaller $\Delta$. For example, set new $\Delta$ value
as $\Delta/\sqrt{10}$, increment the step counter $s=s+1,$ and go to
\textbf{Step 1}.\newline\textbf{Step 4:} Reconstruction with the required
precision is obtained in $p$ iterations or when the maximal allowed number of
iterations $P_{\max}$ is reached. The reconstructed signal is $\hat
{q}(m,n)=y(m,n)=y^{(p)}(m,n)$.

\noindent\

By performing presented iterative procedure, the missing values will converge
to the true signal values, producing the minimal sparsity measure in the ISAR
image domain.

\smallskip

\textbf{Comments on the algorithm:}

- Inputs to the algorithm are the signal size $M\times N$, set of available
signal samples $\mathbf{N}_{A}$, available signal values $q(m_{i},n_{i})$,
$n_{i}\in$ $\mathbf{N}_{A}(m)$, the maximal allowed number of iterations
$P_{\max}$ and the required precision used in \textbf{Step 3}. The algorithm
output is the reconstructed signal matrix $q(m,n)=y(m,n)$.

- When we approach to the optimal point, the gradient algorithm using the
norm-one and a large number of variables (missing signal values) will produce
a solution close to the exact signal samples, with a precision related to the
algorithm step $\Delta$. The precision is improved by using adaptive step
$\Delta$. A value of $\Delta$ equal to the signal magnitude (\ref{deltaINIt})
is used in the starting iteration. When the optimal point is reached then, due
to the norm-one like form, the algorithm will not improve the reconstruction
precision any more, for a given algorithm step $\Delta$. When this case (in
Step 3) is detected the step $\Delta$ is reduced, and the same calculation
procedure is continued from the reached reconstructed signal values. In
several steps, the algorithm can approach the true signal values with a
required precision.

\bigskip

\begin{figure}
[ptb]
\begin{center}
\includegraphics[
height=2.3288in,
width=3.0452in
]%
{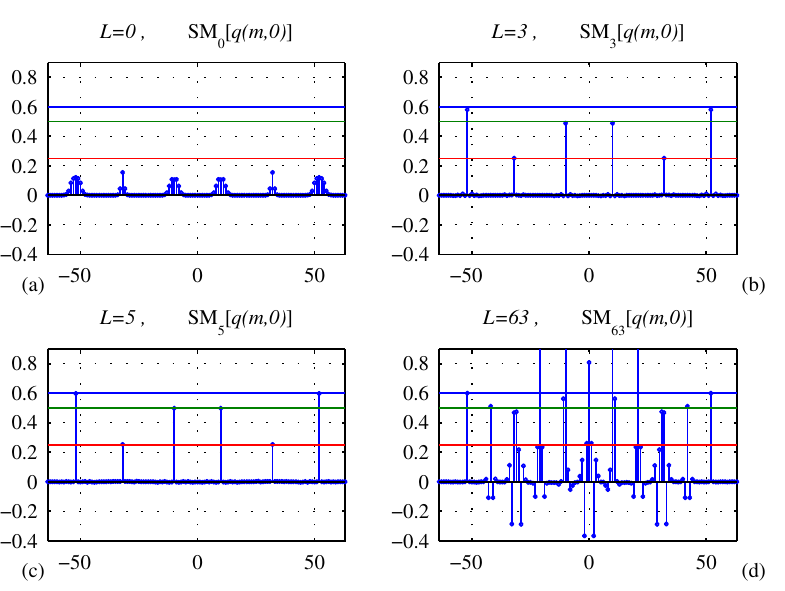}%
\caption{All signal samples (chirps) available: (a) The Fourier transform
based presentation. (b) The S-method with three correction terms, $L=3$. (c)
The S-method with five correction terms, $L=5$ . (d) The Wigner distribution
based presentation (the S-method with $L=63)$. Horizontal lines (red, green,
blue) present the level of the true squared amplitudes of the components. }%
\label{sm_var_l}%
\end{center}
\end{figure}

\noindent\textbf{\textit{{Example 3:} }}A signal corresponding to the Doppler
part of radar signal only is considered first. Its form is
\begin{align*}
q(m,0)  &  =\sum_{i=1}^{6}\sigma_{i}e^{j2\pi\beta_{i}m/M}e^{j\alpha_{i}%
m^{2}/2}\\
&  =2\sqrt{0.6}\cos(52m\pi/64-2.2\pi(m/64)^{2})\\
&  +2\sqrt{1/2}\cos(10m\pi/64+2\pi(m/64)^{2})\\
&  +2\sqrt{1/4}\cos(32\pi m/64-0.75\pi(m/64)^{2})
\end{align*}
with $-64\leq m\leq63$ and $\sigma_{i}\in\{\sqrt{0.6}$, $\sqrt{0.6}$,
$\sqrt{0.5}$, $\sqrt{0.5}$, $\sqrt{0.25}$, $\sqrt{0.25}\}$, $\beta_{i}\in
\{26$, $-26$, $5$, $-5$, $16$, $-16\}$ and $\alpha_{i}\in\{-1.1\pi/1024$,
$1.1\pi/1024$, $\pi/1024$, $-\pi/1024$, $-0.375\pi/1024$, $0.375\pi/1024\}$,
for $i=1,2,...,6$. \ The representations with all available samples are
presented in Fig.\ref{sm_var_l}. The Fourier transform base presentation
(radar image) is shown in Fig.\ref{sm_var_l}(a) for $L=0$ since $SM_{0}%
(k,0)=\left\vert Q(k,0)\right\vert ^{2}$. We can see that although there are
just $6$ reflecting points the number of nonzero (significant) values in
$SM_{0}(k)$ is above $40$. The sparsity condition is heavily degraded. Adding
just a few of the correction terms, according to (\ref{SM_CORR}), and
calculating the S-method based presentation the sparsity in ISAR image is
restored. Presentations with $L=3$ and $L=5$ in the S-method are shown in
Fig.\ref{sm_var_l}(b)-(c). Note that the Wigner distribution, $WD(k,0)=SM_{63}%
(k,0)$, although well concentrated for the components, can not be used due to
emphatic cross-terms which degrade the sparsity, Fig.\ref{sm_var_l}(d).

\begin{figure}
[ptb]
\begin{center}
\includegraphics[
height=2.3289in,
width=3.045in
]%
{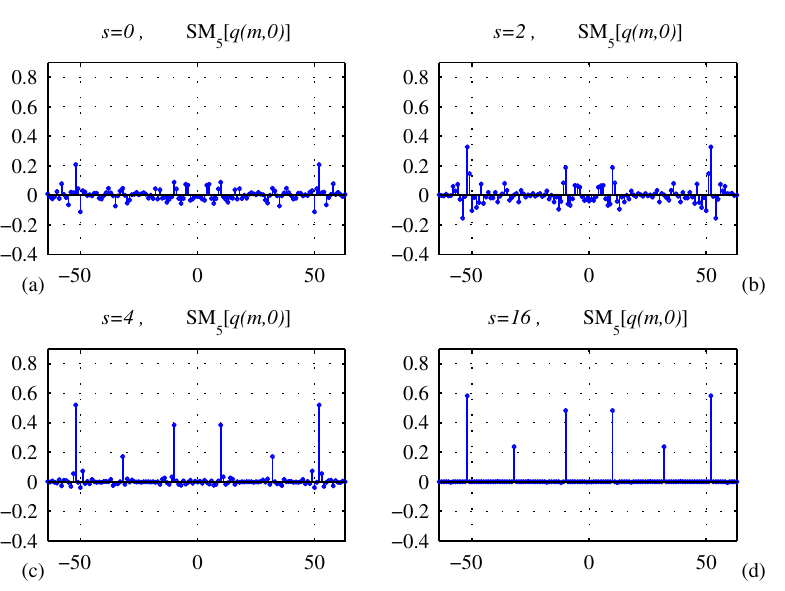}%
\caption{The S-method presentation (radar image) with missing/corrupted 1/3 of
the signal samples (chirps): (a) Initial S-method personation $p=0,$ and the
reconstructed S-method in the next iterations (b)-(d) with $p=2$, $p=4$, and
$p=16$, respectively.}%
\label{sm_var_rec}%
\end{center}
\end{figure}

Consider next the signal with $45$ missing signal values (missing chirps in
this case). Here, the S-method is calculated with $L=5$ and the gradient based
reconstructed algorithm is applied. The S-method, assuming all missing values
are set to $0$, is presented for the initial iteration in Fig.\ref{sm_var_rec}%
(a). The next iterations steps according to the presented iterative algorithm
(denoted by step counter $s$), improve the presentation toward the case as if
all data were available, Fig.\ref{sm_var_rec}(b)-(d) for $s=2,4,$ and $16$.%

\medskip

\begin{figure}
[ptb]
\begin{center}
\includegraphics[
height=3.3092in,
width=2.9564in
]%
{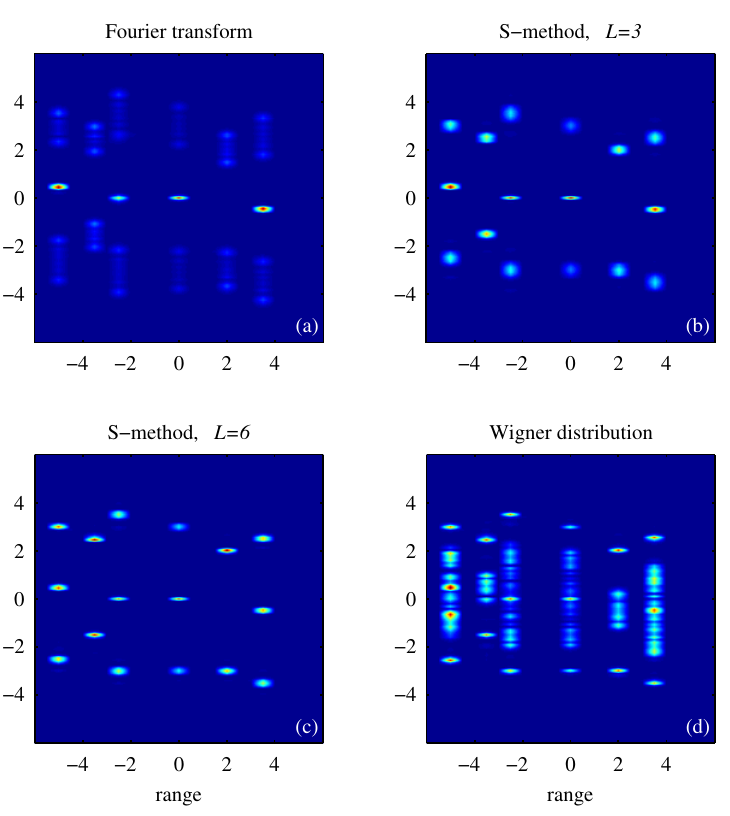}%
\caption{The ISAR image based on: (a) The two-dimensional Fourier transform.
(b) The S-method with $L=3$. (c) The S-method with $L=6$. (d) The Wigner
distribution (the S-method with $L=64$). All data are available.}%
\label{figisarsim_cs_all}%
\end{center}
\end{figure}

\noindent\textbf{\textit{{Example 4:}}} The setup that will be considered for
this ISAR example assumes \cite{Thaya}: a high-resolution radar operating at
the frequency $f_{0}=10.1\,\mathrm{GHz}$, $\Omega_{0}=2\pi f_{0}$, bandwidth
of linear frequency-modulated chirps $B=300\,\mathrm{MHz}$, and the coherent
integration time $T_{c}=2\,\mathrm{s}$. The pulse repetition time is
$T_{r}=T_{c}/256$ with the sampling interval $T_{s}=T_{r}/64$. The target is
at $2\,\mathrm{km}$ distance from the radar, and rotates at $\Omega_{R}%
=4\pi/180\,\mathrm{1/s}=4\,^{\mathrm{o}}\mathrm{/s}$. The nonlinear rotation
with frequency $\Omega=\pi\,\mathrm{1/s}$ is superimposed, $\Omega
_{R}(t)=\Omega_{R}+A\sin(\Omega t),$ and amplitude $A=1.25\pi
/180\,\mathrm{1/s}$ corresponds to the total change in angular frequency
$\Omega_{R}$ for $2.5\pi/180\,\mathrm{1/s}$. Note that here the range and the
cross-range resolutions are $R_{\mathrm{range}}=c/(2B)=0.5\,\mathrm{m}$, and
$R_{\mathrm{cross-range}}=\pi c/(\Omega_{0}T_{c}\Omega_{R})=0.106\,\mathrm{m}$
(calculated for $T_{c}=2\,\mathrm{s}$ with $\Omega_{R}\cong4\pi
/180\,\mathrm{1/s}$, neglecting effects of the nonlinear rotation). It has
been assumed that there are $15$ reflecting points at the positions
$(x_{i},y_{i})\in\{(-3.5,-3.5)$, $(-3.5,-0.5)$, $(-3.5,2.5)$, $(0,-3)$,
$(0,0)$, $(0,3)$, $(2.5,-3)$, $(2.5,0)$, $(2.5,3.5)$, $(3.5,-1.5)$,
$(3.5,2.5)$, $(-2,2)$, $(-2,-3)$, $(5,0.5)$, $(5,3)\}$. First, the case with
all available data is considered. The ISAR image based on the two-dimensional
Fourier transform is presented in Fig.\ref{figisarsim_cs_all}(a). The S-method
based ISAR image with $L=3$ and $L=6$ is shown in in
Fig.\ref{figisarsim_cs_all}(b)-(c). It can be seen that just a few correction
terms to the Fourier transform based ISAR image significantly improve the
concentration. The Wigner distribution (the S-method with $L=64$) is highly
concentrated. However it suffers from the cross-terms,
Fig.\ref{figisarsim_cs_all}(d). The range and cross-range coordinate axes are
scaled with the resolution parameters.%

\begin{figure}
[ptb]
\begin{center}
\includegraphics[
height=3.3092in,
width=2.9564in
]%
{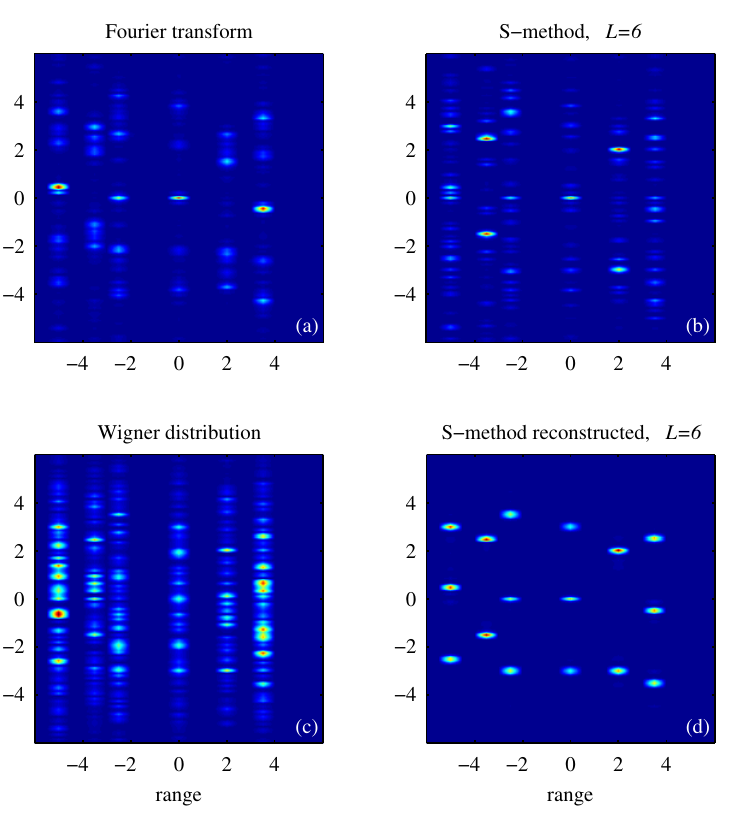}%
\caption{The ISAR image based on: (a) The two-dimensional Fourier transform.
(b) The S-method with $L=6$. (c) The Wigner distribution. (d) The S-method
based on the reconstructed signal in two steps ($s=2)$. Only 50\% of randomly
positioned available data are used in the reconstruction.}%
\label{figisarsim_cs_rec}%
\end{center}
\end{figure}

The case with $50\%$ of the data being unavailable (or removed due heavy
corruption) is considered next. The ISAR image calculated by using the
two-dimensional Fourier transform is presented in Fig.\ref{figisarsim_cs_rec}%
(a). As we can see the image sparsity is low. Since a large amount of the data
is missing this image can not be improved by a direct application of the
S-method since the missing data behave as a noise (Section II.A). The S-method
based image with $L=6$ is shown in Fig.\ref{figisarsim_cs_rec}(b). The same
holds for the Wigner distribution (the S-method with $L=64$) given in
Fig.\ref{figisarsim_cs_rec}(c). However, the original image calculated with
the S-method is highly sparse. Therefore the unavailable data can be
reconstructed by minimizing the S-method subject to the available data,
equation (\ref{DEF_L1}). The gradient based method is used to solve this
minimization problem. The reconstructed S-method is almost the same as the
S-method of the signal with all available data. It is presented in
Fig.\ref{figisarsim_cs_rec}(d).%

\section{Conclusion}

An analysis of the ISAR image reconstruction in the case of a large number of
unavailable or heavily corrupted data is presented. A simple method that can
produce reconstruction in the case of uniform motion is reviewed, along with a
simple an accurate analysis of the noise influence to the results. In the case
of fast and complex target manoeuvring the ISAR image is blurred and the
sparsity property is lost. For a large number of reflecting points a
parametric approach to refocus the image and reconstruct the signal with large
number of missing data would be computationally extensive. A simple
nonparametric method is used here to refocus image. Since it belongs to the
class of quadratic time-frequency representations, a direct linear relation
between the sparsity domain and the signal can not be established. Thus, the
reconstruction task is appropriately reformulated. An adapted form of gradient
algorithm is used is recover the ISAR image of the quality as in the case if
all data were available. The efficiency of the proposed methods is illustrated
on several numerical examples.

\newpage

\begin{IEEEbiography}[{\includegraphics[width=1in,height=1.25in,clip,keepaspectratio]{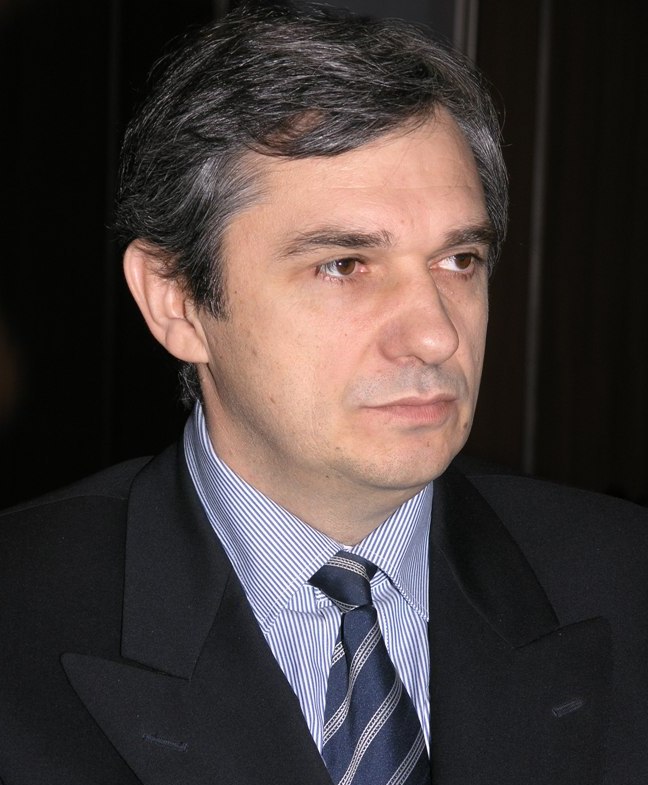}}]%
{Ljubi\v{s}a Stankovi\'{c}}
 (Fellow IEEE) was born in
Montenegro on June 1, 1960. He received the B.S. degree from the University of
Montenegro, Podgorica, in 1982, the M.S. degree from the University of
Belgrade, Belgrade, Yugoslavia, in 1984, and the Ph.D. degree from the
University of Montenegro in 1988, all in electrical engineering. In 1982 he
received the "Best Student at the University" award from the University of
Montenegro. As a Fulbright grantee, he spent the 1984-1985 academic year at
the Worcester Polytechnic Institute, Worcester, MA. Since 1982, he has been on
the faculty at the University of Montenegro, where he has been a full
professor since 1995. In 1997-1998 and 1999, he was on leave at the Ruhr
University Bochum, Bochum, Germany, with Signal Theory Group, supported by the
Alexander von Humboldt Foundation. At the beginning of 2001, he spent a period
of time at the Technische Universiteit Eindhoven, Eindhoven, The Netherlands,
as a visiting professor. During the period of 2003-2008, he was Rector of the
University of Montenegro. He was also active in politics, as a vice president
of the Republic of Montenegro from 1989 to 1991 and a member of the Federal
Parliament of Yugoslavia 1992-1996. Prof. Stankovic is the ambassador of
Montenegro to the United Kingdom, Republic of Ireland, and Iceland. He is also
a vistiting academic at the Imperial College, London. His current interests
are in time-frequency analysis, radar signal processing, and sparse signal
processing. He published more than 350 technical papers, more than 130 of them
in the leading international journals, mainly the IEEE editions. He published
a book "\textit{Time-Frequency Signal Analysis with Applications}", with
Artech House in 2013. He authored several chapters in reference books and
monographs published by Academic Press, Elsevier, and CRC. Prof. Stankovi\'{c}
received the highest state award of the Republic of Montenegro in 1997, for
scientific achievements. His group received a research award grant in 2001
from the Volkswagen Foundation, Germany. He was a member the IEEE Signal
Processing Society's Technical Committee on Theory and Methods, and an
Associate Editor of the \textit{IEEE Transactions on Image Processing},
\textit{IEEE Signal Processing Letters}, and \textit{IEEE Transactions on
Signal Processing}. Prof. Stankovic is a member of the National Academy of
Science and Art of Montenegro (CANU) since 1996 and a member of the European
Academy of Sciences and Arts.
\end{IEEEbiography}

\end{document}